\documentclass{ws-procs9x6} 
 
\begin{document} 
\title{Neutrino Experiments and Their Implications} 
 
\author{A.~B. BALANTEKIN}

\address{University of Wisconsin, Department of Physics \\ 
Madison, WI  53706,  USA \\ 
E-mail: baha@nucth.physics.wisc.edu} 
 
\maketitle 

\abstracts{Recent developments in solar, reactor, and accelerator 
neutrino physics are reviewed. Implications for neutrino physics, 
solar physics, nuclear two-body physics, and r-process nucleosynthesis 
are briefly discussed.} 

\section{Introduction}

\begin{figure} 
\includegraphics[scale=0.35]{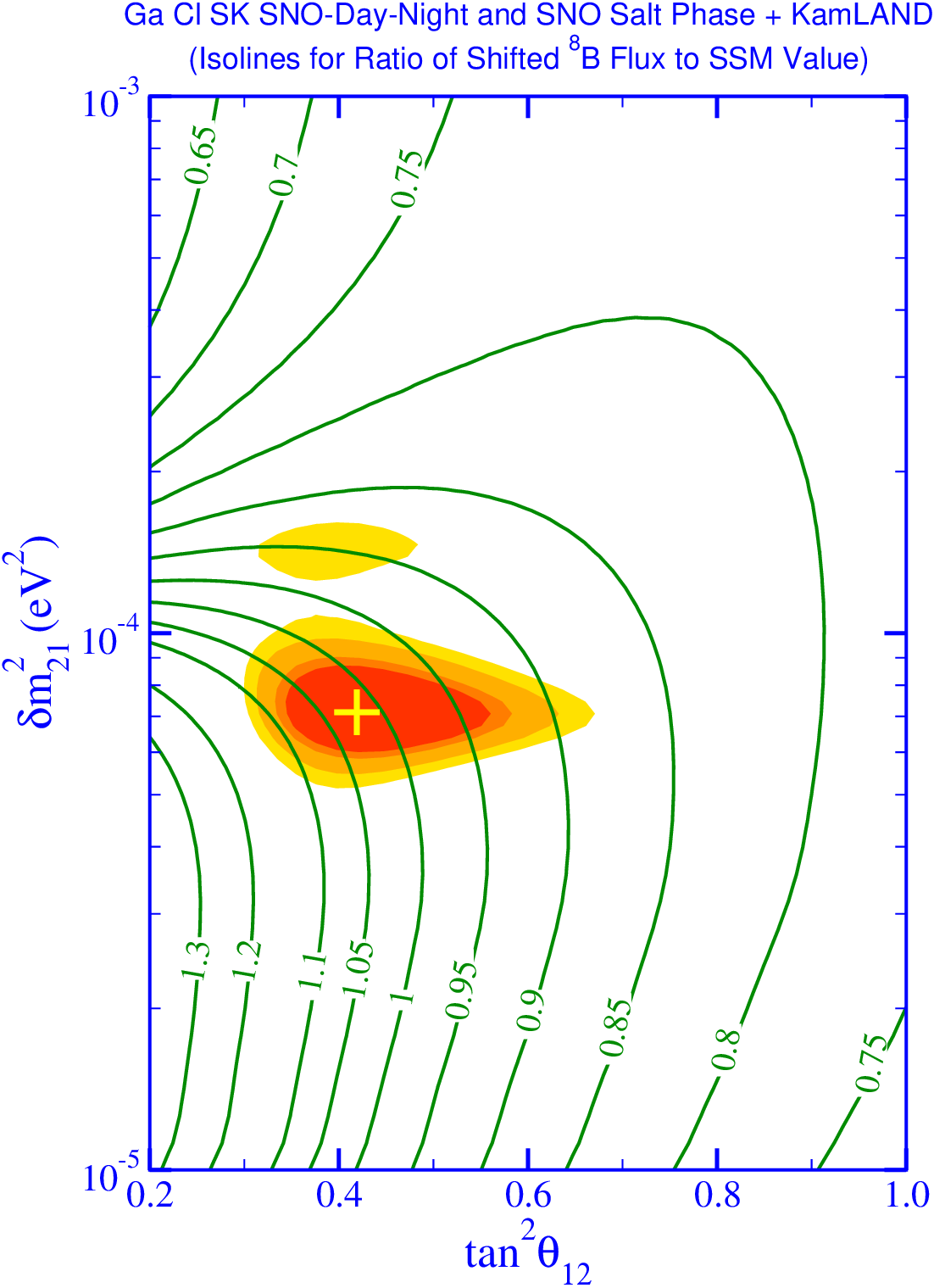} 
\vspace*{0cm} \caption{ \label{fig:1} 
Allowed confidence levels from the joint analysis of all 
available solar neutrino data (chlorine, average gallium, SNO and SK 
spectra and SNO salt phase) and KamLAND reactor data 
The isolines are the ratio of the shifted $^8$B flux 
to the SSM value. 
At best fit (marked by a cross) the value of this ratio is determined 
to be $1.02$ (from Reference 10).} \end{figure}

Solar neutrino experiments, especially with the announcement of recent 
results 
from the Sudbury Neutrino Observatory (SNO) \cite{Ahmed:2003kj}, have 
reached the precision stage.  
An analysis of the data from SNO as well as data from other solar 
neutrino experiments (Super-Kamiokande [SK] 
\cite{Fukuda:2002pe}, Chlorine \cite{Cleveland:nv}, and 
Gallium \cite{Abdurashitov:2002nt,Hampel:1998xg,Altmann:2000ft}), 
combined with the data from the reactor experiment KAMLAND 
\cite{Eguchi:2002dm}, place severe 
constraints on the neutrino parameters, especially mixing between first and 
second generations 
\cite{Balantekin:2003dc,deHolanda:2003nj,Balantekin:2003jm}. 
The neutrino parameter space obtained from such a  global analysis, 
including the neutral-current results from the SNO salt phase, 
is shown in Fig. \ref{fig:1} \cite{Balantekin:2003jm}. 
 
The mixing angle between first and second generations of the neutrinos 
dominates the solar neutrino oscillations whereas the mixing angle between 
second and third generations dominates the oscillations of atmospheric 
neutrinos. 
There are several puzzles in the data. Both mixing angles seem to be close 
to maximum, very unlike the mixing between quarks. Also the third mixing 
angle, between first and third generations, seems to be very small, even 
possibly zero. 
It is especially important to find out if this mixing angle is indeed 
different from zero since in the mixing matrix it multiplies a 
CP-violating phase. Such a CP-violation may have far reaching 
consequences. 
To explain the baryon excess (over antibaryons) in  
the Universe, Sakharov pointed out that it may be sufficient to satisfy  
three conditions: i) Baryon number non-conservation (which is readily  
satisfied by the grand unified theories), ii) CP-violation, and iii) 
Non-equilibrium conditions. It is entirely possible that the CP-violation  
necessary for the baryogenesis is hidden in the neutrino sector. 
 
\begin{figure}[t] 
\includegraphics[scale=0.29]{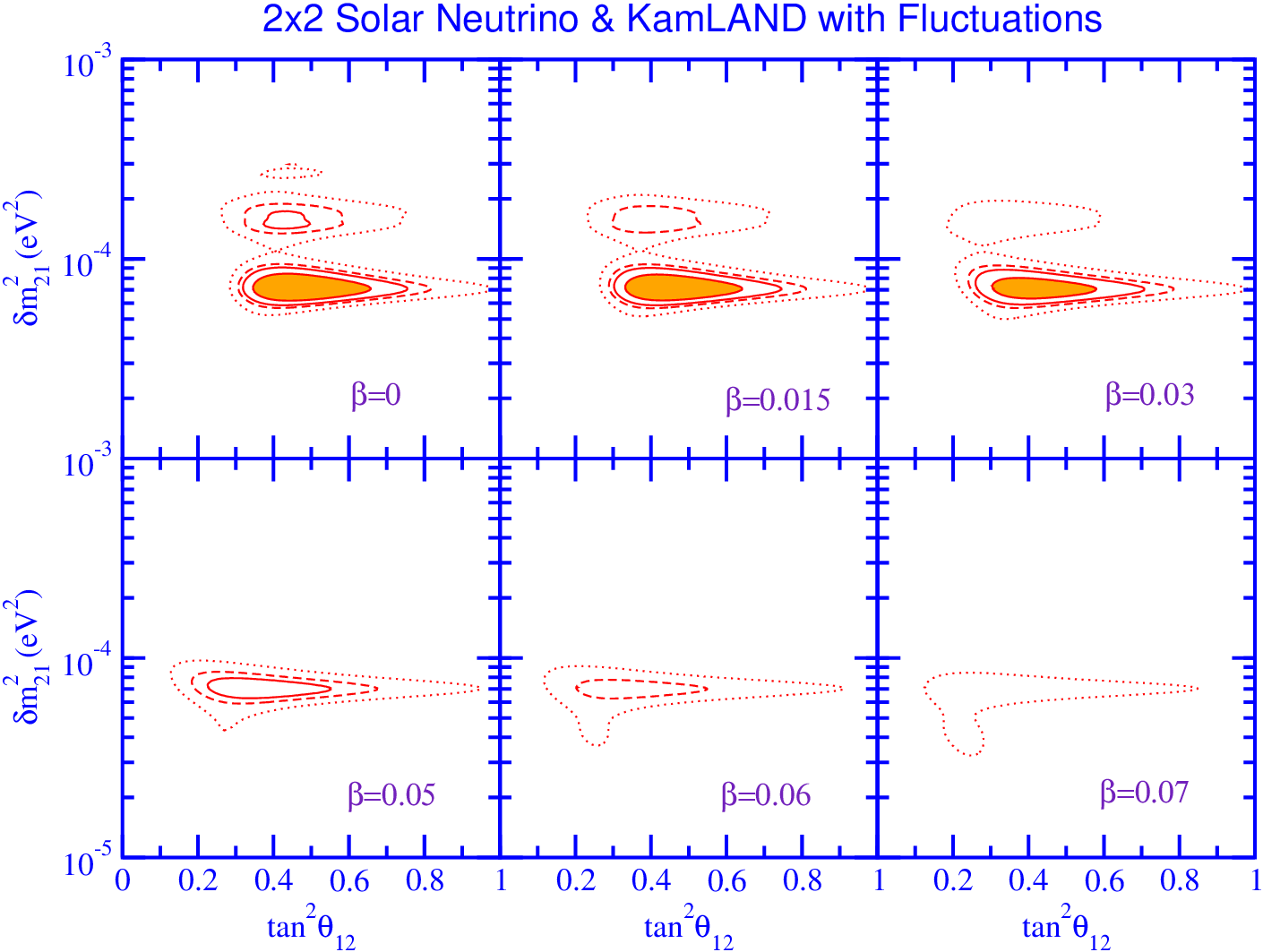}
\vspace*{0cm} \caption{ \label{fig:2}  Allowed regions of the 
  neutrino parameter space with solar-density fluctuations when the 
  data from the solar neutrino and KamLAND experiments are used. 
  The SSM density profile of 
  Reference 14 and the correlation length of 10 km are 
  used. The case with no fluctuations ($\beta=0$) are compared with 
  results obtained with the indicated fractional fluctuation.  The 
  shaded area is the 70 \% confidence level region. 90 \% (solid 
  line), 95 \% (dashed line), and 99 \% (dotted line) confidence 
  levels are also shown (From Reference 15).}  
\end{figure} 

It is worth pointing out that  
high-precision solar-neutrino data have potential beyond exploring neutrino 
parameter space. Here we discuss two such applications to solar physics and 
to nuclear physics. 

\section{Limits on Solar Density Fluctuations}

It was suggested that solar neutrino data can be inverted 
to extract information about the density scale height 
\cite{Balantekin:1997fr} in a similar way the helioseismological 
information is inverted to obtain the speed of the sound 
throughout the Sun. Even though the precision of the data has not yet 
reached to a point where such an inversion is possible, one can obtain 
rather strong limits on {\em fluctuations} of the solar density using the 
current solar neutrino data. 
To do so one assumes \cite{Loreti:1994ry} that 
the electron density $N_e$ fluctuates around the 
value, $\langle N_e \rangle$, predicted by the Standard Solar Model (SSM) 
\cite{Bahcall:2000nu} 
\begin{equation} 
N_e (r) = (1 + \beta F (r)) \langle N_e (r) \rangle , 
\label{flucdef}
\end{equation} 
and that the fluctuation $F (r)$ takes the form of 
white-noise. It turns out 
that the effect of the fluctuations is more dominant when
the neutrino parameters and the average density are such that neutrino
evolution in the absence of fluctuations is adiabatic.
There are two constraints on
the value of the correlation length. 
One is a restriction in the applicability of our analysis. 
In averaging over the
fluctuations we assumed that the correlation function is a delta
function. In the Sun it is more physical to
imagine that the correlation function is like a step function of size
$\tau$. Assuming that the logarithmic derivative is small, which is
accurate for the Sun, delta-correlations are approximately the same as
step-function correlations if the condition
\begin{equation}
\tau \ll \left( \sin 2\theta \frac{\delta m^2}{2E} \right)^{-1}
\label{eq:step-approx-msw}
\end{equation}
is satisfied \cite{Balantekin:1996pp}. 
A second constraint on the correlation length is provided by the
helioseismology. Density fluctuations over scales of $\sim 1000$ km
seem to be ruled out. 
On the other hand current helioseismic observations are rather
insensitive to density variations on scales close to $\sim 100$ km
\cite{Burgess:2002we}.

The neutrino parameter space 
for various values of the parameter $\beta$ was calculated in Reference  
15 and is shown in Figure 2. These results, in 
agreement with the calculations of other authors 
\cite{Burgess:2003su,Guzzo:2003xk}, show that the neutrino data constrains 
solar density fluctuations to be less than $\beta = 0.05$ at the 70 \% 
confidence level when $\tau$ is about 10 km. 
It is important to emphasize that the best fit 
to the combined solar neutrino and KamLAND data is given by $\beta = 
0$ (exact SSM). 
Neutrinos interact with dense matter 
not only in the Sun (and other stars) but also 
in several other sites such as
the early universe, supernovae, and newly-born neutron stars and
neutrino interactions with a stochastic background may play an even
more interesting role in those sites. 

\section{Two-Body Axial Current}

In the effective field theory approach to nuclear interactions,  
nonlocal interactions at 
short distances are represented by effective local interactions 
in a derivative expansion. Since the effect of a given operator on 
low-energy physics is inversely proportional to its dimension, 
an effective theory valid at low energies can be 
written down by retaining operators up to a given dimension. 
It turns out that the deuteron break-up reactions 
\begin{equation} 
\nu_ e + d \rightarrow e^- +  p + p
\end{equation}
and 
\begin{equation}
\nu_x + d \rightarrow \nu_x + p + n, 
\end{equation}
observed at SNO, are dominated by a $^3S_1 \rightarrow ^3S_0$ 
transition, hence one only needs the coefficient of the two-body 
counter term, commonly called $ L_{1A} $, to parameterize the unknown 
isovector axial two-body current \cite{Butler:1999sv}. 
Chen, Heeger, and Robertson, using the SNO  
and SK charged-current, neutral current, and 
elastic scattering rate data, 
found \cite{Chen:2002pv} $L_{1A}  = 4.0 \pm 6.3 \: \mathrm{fm}^3$. 
In order to obtain this result they wrote the observed 
rate in terms of an averaged effective cross section and a 
suitably defined response function. One can explore 
the phenomenology associated with the variation of $L_{1A} $. For example 
the variation of the neutrino parameter space, which fits the SNO data, as 
$L_{1A} $ changes 
was calculated in \cite{Balantekin:2003ep} and is shown 
in Figure 3. In Reference 21 the 
most conservative fit value with fewest assumptions is found to be 
$ L_{1A} = 4.5 ^{+18}_{-12} \: \mathrm{fm}^3$. 
(One should point out that if the neutrino parameters were better 
known one can get a much tighter limit).   
It was also shown that the contribution of the uncertainty of 
$ L_{1A} $ to 
the analysis and interpretation of the solar neutrino data measured 
at the Sudbury Neutrino Observatory is significantly less than the 
uncertainty coming from the lack of having a better knowledge 
of $ \theta_{13} $, the mixing angle between first and third generations. 

\begin{figure}[t]  
\includegraphics[scale=0.3]{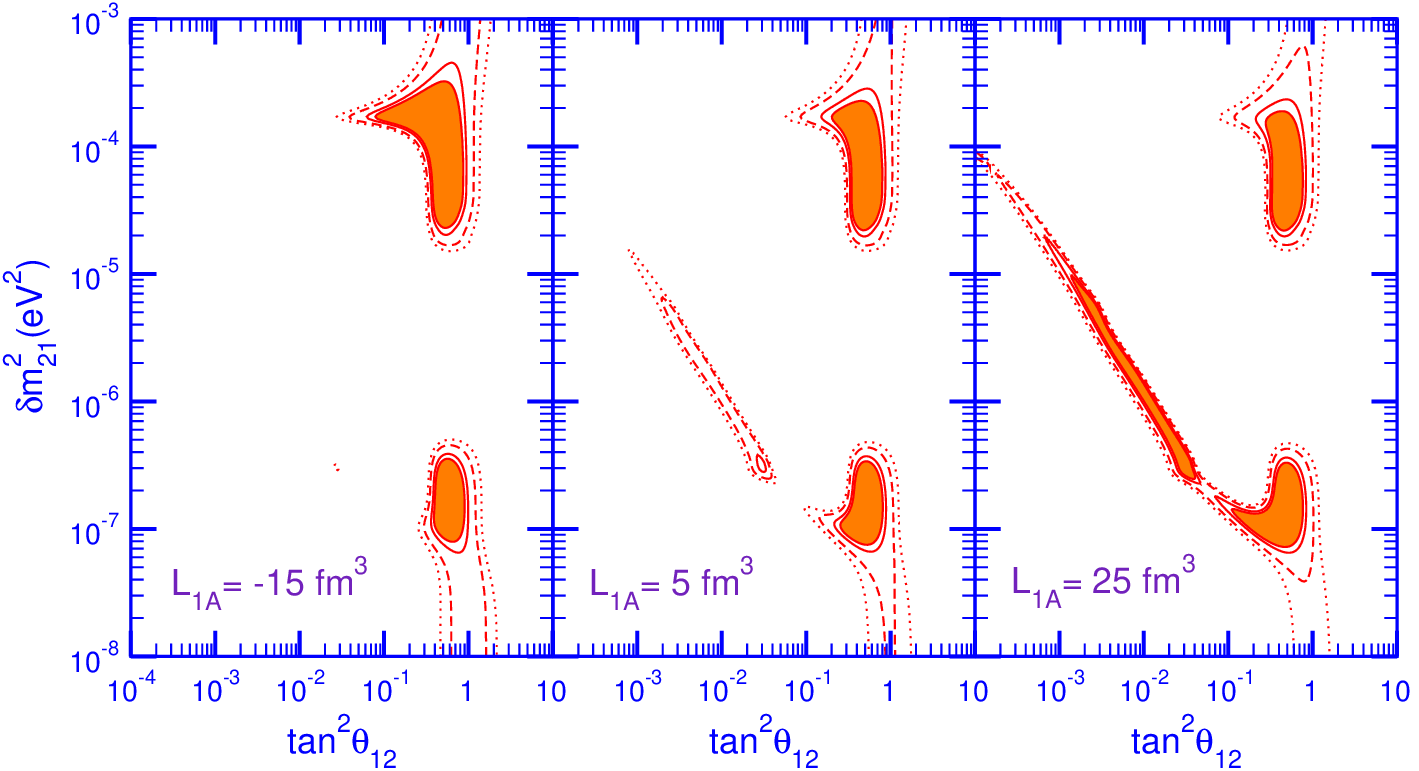}  
\vspace*{0cm} \caption{ \label{fig:3}  
The change in the allowed region of the neutrino parameter space  
using solar neutrino data measured at SNO as the value of  
$ L_{1A} $ changes. The shaded  
areas are the 90 \% confidence level region. 95 \% (solid line),  
99 \% (log-dashed line), and 99.73 \% (dotted-line) confidence  
levels are also shown (From Reference 21).}  
\end{figure}  

\section{Implications for r-process Nucleosynthesis}

There is another puzzling experimental result. 
The Los Alamos Liquid Scintillator Neutrino Detection (LSND)
experiment has reported an excess of $\bar\nu_e$-induced events above
known backgrounds in a $\bar\nu_\mu$ beam with a statistical
significance of $3$ to $4$ $\sigma$
\cite{Athanassopoulos:1996jb,Aguilar:2001ty}. 
The mass scale indicated by the LSND experiment is drastically 
different than the mass scales implied by the solar and atmospheric 
neutrino experiments. Since to get three different differences one needs 
four numbers, a confirmation of the LSND result by the mini-BooNE 
experiment represents evidence for vacuum neutrino oscillation at a new
$\delta m^2$ scale.  Discovery of such a mixing
would imply either CPT-violation  in the
neutrino sector, or the existence of a light singlet sterile 
neutrino which mixes with active species. 
The latter explanation may
signal the presence of a large and unexpected net lepton number in the
universe. The existence of a light singlet complicates the
extraction of a neutrino mass limit from Large Scale Structure data.
It may also have implications for core-collapse supernovae, which is one 
of the leading candidates for the site of r-process nucleosynthesis 
\cite{Balantekin:2003ip}. 
A sterile neutrino scale implied by the LSND experiment may resolve some 
outstanding problems preventing a successful nucleosynthesis. 
Formation of too many alpha particles
in the presence of a strong electron neutrino flux coming from the cooling of 
the proto-neutron star, known as the
alpha effect \cite{Fuller:ih,Meyer:1998sn}, may be prevented by 
transforming active electron 
neutrinos into sterile neutrinos 
\cite{McLaughlin:1999pd,Caldwell:1999zk,Patel:1999hm,fetter}. One can find 
the appropriate mass scale to achieve this goal 
\cite{McLaughlin:1999pd,fetter} which seems to overlap with the LSND mass 
scale.

R-process nucleosynthesis requires a neutron-rich environment, i.e.,
the ratio of electrons to baryons, $Y_e$, should be less than one
half. Time-scale arguments based on meteoritic data suggests that one
possible site for r-process nucleosynthesis is the neutron-rich
material associated with core-collapse supernovae
\cite{Qian:1998,Qian:1998cz}. In one model for neutron-rich material
ejection following the core-collapse, the material is heated with
neutrinos to form a ``neutrino-driven wind''
\cite{Woosley:ux,janka}. In outflow models freeze-out from nuclear
statistical equilibrium leads to the r-process nucleosynthesis. The
outcome of the freeze-out process in turn is determined by the
neutron-to-seed ratio. The neutron to seed
ratio is controlled by the expansion rate, the neutron-to-proton 
ratio, and the entropy per baryon. Of these the neutron-to-proton ratio 
is controlled by the flavor composition of the 
neutrino flux coming from the cooling of the 
proto-neutron star. Hence understanding neutrino properties (including 
the impact of neutrino-neutrino scattering in neutrino propagation 
\cite{Qian:wh}) could  
significantly effect our understanding of the r-process nucleosynthesis. 

I thank G. Fuller, G. McLaughlin, and H. Y\"uksel for many useful 
discussions and the organizers of the OMEG03 conference for their 
hospitality. This work was supported in part by the U.S. National Science 
Foundation Grant No.\ PHY-0244384 and in part by 
the University of Wisconsin Research Committee with funds granted by 
the Wisconsin Alumni Research Foundation.

\end{document}